\documentclass[referee]{raa}
\usepackage{graphicx,times}
\usepackage{natbib}
\usepackage{amssymb,amsmath}
\bibpunct{(}{)}{;}{a}{}{,}

\usepackage[a4paper=true,dvipdfm=true,pagebackref=true]{hyperref}
\hypersetup{pdftitle = The title of my PDF, pdfauthor = My name, pdfsubject= The subject, pdfkeywords = keyword1 keyword2 keyword3}
\hypersetup{colorlinks = true, linkcolor = green, anchorcolor = red, citecolor = blue, filecolor = red, pagecolor = red, urlcolor = red}

\begin{document}

\title{ What is parameterized $Om(z)$ diagnostics telling us in light of recent observations?}
 \volnopage{ {\bf 2012} Vol.\ {\bf X} No. {\bf XX}, 000--000}
   \setcounter{page}{1}

\author{Jing-Zhao Qi\inst{1}, Shuo Cao\inst{1}, Marek Biesiada\inst{1,2}, Tengpeng Xu\inst{1}, Yan Wu\inst{1}, Sixuan Zhang \inst{1} and Zong-Hong Zhu\inst{1}}

\institute{Department of Astronomy, Beijing Normal University,
Beijing, 100875, China; {\it caoshuo@bnu.edu.cn}\\
\and
Department of Astrophysics and Cosmology, Institute of Physics,
University of Silesia, Uniwersytecka 4, 40-007, Katowice, Poland \\
\vs \no
   {\small Received XX; accepted XX}
}

\abstract{
In this paper, we propose a new parametrization of $Om(z)$
diagnostics and show how the most recent and significantly improved
observations concerning the $H(z)$ and SN Ia measurements can be
used to probe the consistency or tension between $\Lambda$CDM model
and observations. Our results demonstrates that $H_0$ plays a very
important role in the consistency test of $\Lambda$CDM with the
$H(z)$ data. Adopting the Hubble constant priors from
\textit{Planck} 2013 and Riess (2016), one finds a considerable
tension between the current $H(z)$ data and $\Lambda$CDM model and
confirms the conclusions obtained previously by the others. However,
with the Hubble constant prior taken from WMAP9, the discrepancy
between $H(z)$ data and $\Lambda$CDM disappears, i.e., the current
$H(z)$ observations still support the cosmological constant
scenario. This conclusion is also supported by the results derived
from the JLA SNe Ia sample. The best-fit Hubble constant from the
combination of $H(z)$+JLA ($H_0=68.81^{+1.50}_{-1.49}$ km/s/Mpc) is
well consistent with the results derived both by Planck 2013 and
WMAP9, which is significantly different from the recent local
measurement by Riess (2016).
}

   \authorrunning{J.-Z. Qi et al. }            
   \titlerunning{Using parameterized $Om(z)$ diagnostics to test $\Lambda$CDM}  

\maketitle

\section{Introduction} \label{introduction}

The fact that our universe is undergoing an accelerated expansion at
the present stage has become one of the most important issues of the
modern cosmology ever since the indication of it came from observations of type
Ia supernovae (SN Ia) (\citealt{riess1998supernova,perlmutter1999measurements}), which was
also supported by other independent astrophysical observations
including large scale structure (\citealt{tegmark2004cosmological}),
baryon acoustic oscillation (BAO) peaks (\citealt{eisenstein2005detection}) and cosmic microwave background
(CMB) (\citealt{spergel2003wmap}). This phenomenon poses a great mystery concerning what component of our universe could produce a
repulsive force to drive this accelerating expansion.
In the framework of general relativity, a mysterious substance with
negative pressure, dubbed as dark energy, was proposed to explain
this acceleration. Due to still unknown nature of dark energy, the investigation of its equation of state (EoS), $w=p/\rho$, a
critical parameter to characterize the dynamical property of dark
energy, has also become a significant research theme in modern
cosmology. Many cosmologists suspect that dark
energy is just the cosmological constant with $w=-1$, due to its simplicity
and a remarkable consistency with almost all observational
data. However, the notable fine-tuning problem (\citealt{weinberg1989cosmological}) and coincidence problem (\citealt{1999PhRvL..82..896Z}) still question why $\Lambda$CDM is
declared to be the concordance cosmological model to describe the
overall evolution of the Universe. Thereupon, the possibility that
cosmic EoS is a variable depending with time has been explored in a
number of dynamical dark energy models, such as quintessence (\citealt{Caldwell:2005tm,Zlatev:1998tr}), K-essence (\citealt{chiba2000kinetically,armendariz2000dynamical}), phantom (\citealt{kahya2007quantum,onemli2004quantum,singh2003cosmological}),
etc. In the face of so many competing dark energy candidates, it is
important to find an effective way to decide whether the EoS of dark
energy is time varying, which is significant for us to understand
the nature of dark energy.

Following this way, an effective diagnostic named $Om(z)$, which is
sensitive to the EoS of dark energy and thus provides a null test of
$\Lambda$CDM model, was initially introduced by (\citet{sahni2008two})
and extensively studied in many subsequent works (\citealt{sahni2014model,ding2015there,2016ApJ...825...17Z}). If the
value of $Om(z)$ is a constant at any redshift, dark energy is
exactly in the form of the cosmological constant, whereas the
evolving $Om(z)$ corresponds to other dynamical dark energy models.
On the other hand, the slope of $Om(z)$ could
distinguish two different types of dark energy models, i.e., a
positive slope indicates a phase of phantom ($w<-1$) while a
negative slope represents quintessence ($w>-1$) (\citealt{sahni2008two}). Based on the above
results, many previous works have performed consistency tests of the
$\Lambda$CDM model, by using reconstructed $Om(z)$ with the
combination of Gaussian processes and observations including SN Ia
and Hubble parameter data (\citealt{seikel2012using,qi2016testing,yahya2014null}). It was found
that $\Lambda$CDM is compatible with Union2.1 SN Ia and smaller
sample of $H(z)$ measurements. More recently, \citet{shafieloo2012new} developed an improved version of the
two-point diagnostics $Omh^2(z_1,z_2)$, which was also extensively
used to test $\Lambda$CDM with different samples of $H(z)$ data (\citealt{sahni2014model,ding2015there,2016ApJ...825...17Z}). The
general conclusion, which revealed the tension between $H(z)$ data
and $\Lambda$CDM in the framework of Planck data (\citealt{ade2014planck}), implies that the $\Lambda$CDM model may not
be the best scenario of our universe, or dark energy does not exist
in the form of the cosmological constant. Considering the
significance of this result to understand the nature of dark energy,
it is still important to seek its confirmation with alternative techniques.

\begin{figure}
\centering
\includegraphics[width=8cm,height=6cm]{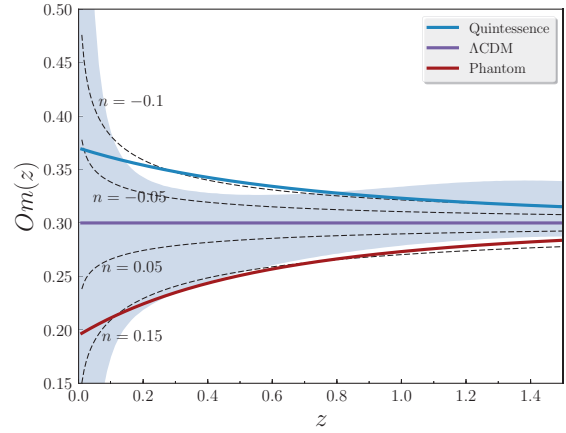}
\caption{The evolution of $Om(z)$ versus the redshift $z$ (black
dashed lines) for the parametrization of $Om(z)$ with $n=-0.1,\
-0.05,\ 0.05,\ 0.15$. Three different cosmologies ($\Lambda$CDM,
quintessence and phantom) denoted by solid lines are also added for
comparison. The light blue shadow area represents the $1\sigma$
confidence region of $Om(z)$ reconstructed by GP-processed $H(z)$
data.}\label{show}
\end{figure}

\begin{figure}
\centering
\includegraphics[width=8cm,height=6cm]{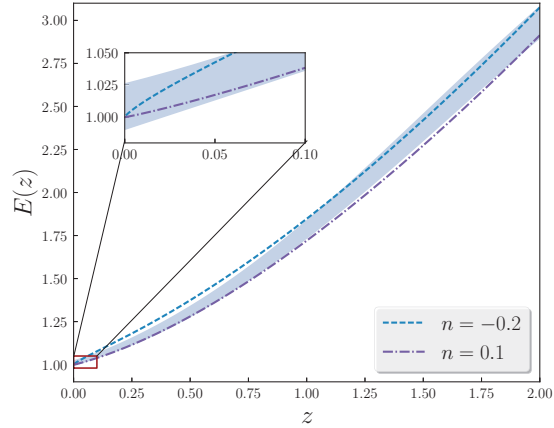}
\caption{The evolution of $E(z)$ versus the redshift $z$ for the
parametrization of $Om(z)$ with $n=-0.2,\ 0.1$. The light blue
shadow area represents the $1\sigma$ confidence region of $E(z)$
reconstructed by GP-processed $H(z)$ data.}\label{show_Ez}
\end{figure}

In this paper, we propose a parametrization of $Om(z)$ to
provide a null test of $\Lambda$CDM, which successfully alleviates
the disadvantages of the traditional $Om(z)$ associated with its strong
dependence on the smoothing data methodology (\citealt{seikel2012using,qi2016testing,yahya2014null}) and the
statistical approach used (\citealt{sahni2014model,ding2015there,2016ApJ...825...17Z}). With this
new parametrization of $Om(z)$, the purpose of this work is to show
how the combination of the most recent and significantly improved
observations regarding the $H(z)$ and SN Ia can be
used to probe the consistency or reveal the tension between the $\Lambda$CDM model
and observations. This paper is organized as follows: In section
\ref{method} we briefly introduce the $Om(z)$ and its newly-proposed
parametrization. In section \ref{results}, we use the latest $H(z)$
data to constrain the $Om(z)$ parameters and make comparison with
the results obtained from Planck, WMAP9 (\citealt{hinshaw2013nine}), and
a local determination of $H_0$ from \citet{riess20162}. Consistency
test of $\Lambda$CDM with JLA SN Ia sample is also shown in
section~\ref{JLA}. Finally, the conclusions are summarized in
section \ref{conclusion}.

\section{Methodology and data} \label{method}

Considering the flat Friedmann-Lema\^{i}tre-Robertson-Walker
spacetime, the general Friedmann equation for the Universe filled with a
perfect fluid with an equation of state $w(z)$ (in addition to
pressureless matter and now dynamically negligible radiation) can be
written as
\begin{eqnarray}
E^2(z)&\equiv & \frac{H^2(z)}{H^2_0} = \Omega_{m0}(1+z)^3+(1-\Omega_{m0}) \nonumber \\
&&\times \exp \left(3\int_0^z \frac{1+w(z')}{1+z'}dz' \right), \label{fried}
\end{eqnarray}
where $\Omega_{m0}$ is the present matter density of the Universe.
Inspired by the form of this equation in the $\Lambda$CDM case, the $Om(z)$
diagnostic has been defined as (\citealt{sahni2008two})
\begin{equation}
Om(z) \equiv \frac{E^2(z)-1}{(1+z)^3-1}. \label{om}
\end{equation}
It is obvious that in the flat $\Lambda$CDM model, the $Om(z)$ evaluated at
any redshift is
always equal to the present mass density parameter $\Omega_{m0}$.

Therefore, from the observations of the expansion rates at different
redshifts, we would be able to differentiate between $\Lambda$CDM
and other dark energy models including evolving dark energy. For
instance, for the simplest phenomenology of dark energy with
constant equation of state parameter $w=const$, a positive slope of
$Om(z)$ relates to a phase phantom ($w<-1$) and a negative slope
represents the quintessence model ($w>-1$) (\citealt{sahni2008two}),
which is shown in Fig.~\ref{show}. Motivated by the physical
indication of $Om(z)$ slope and the well-known
Chevalier-Polarski-Linder (CPL) model concerning reconstruction of
evolving dark energy EoS, we propose the following theoretical
parametrization for $Om(z)$:
\begin{equation}
Om(z)=\alpha \left(\frac{z}{1+z}\right)^n,
\end{equation}
where $\alpha$ and $n$ are the two constant parameters. From the
above expression, it is straightforward to show that $\Lambda$CDM is
fully recovered when $\alpha=\Omega_{m0}$ and $n=0$. Moreover, from
a simple comparison illustrated in Fig.~\ref{show}, one may easily
find a positive slope $n>0$ indicates a phase of phantom, while a
negative slope represents quintessence-like models. Compared with
the direct study on the equation of state of dark energy in the
previous works \citep{cao2014cosmic}, the introduction of the new
parameter $n$ provides a new cosmological-model-independent method
to differentiate a wider range of cosmological solutions with
effective equation of state, which focus on gravitational
modifications (i.e., $f(R)$ and $f(T)$ gravity) to account for the
cosmic acceleration without the inclusion of exotic dark energy
\citep{chiba20031,wu2011f,Qi:2017xzl}. We remark that one
disadvantage of this $Om(z)$ parameterization is that it would be
divergent at $z=0$ when $n<0$. However, as is shown in
Fig.~\ref{show}, the $Om(z)$ reconstructed by $H(z)$ data (see Table
\ref{tab: Hz}) with Gaussian processes (GP), which is consistent
with this parametrization within $1 \sigma$ confidence level,
exhibits the similar divergence feature at $z\sim 0$. Another
disadvantage of this parameterization in this analysis lies in the
strong assumption that the slope parameter $n$ is a constant, which
only proposes a special candidate to test the possible crossing of
the cosmological constant boundary with different value of $n$. In
order to make a comparison with other cosmological models including
quintom cosmology \citep{Cai:2009zp} and other modified gravity
models \citep{chiba20031,wu2011f,Qi:2017xzl}, a possible solution is
to generalize the slope parameter $n$ as a function of redshift $z$,
which will be considered in our future work concentrating on more
cosmological applications. Now, the dimensionless Hubble parameter
can be rewritten as
\begin{eqnarray}
E^2(z)&=&Om(z)\left[(1+z)^3-1\right]+1 \nonumber \\
&=&\alpha \left(\frac{z}{1+z}\right)^n\left[(1+z)^3-1\right]+1,
\label{Eom}
\end{eqnarray}
and further used to estimate the values of $\alpha$ and $n$ from
various observational data by minimizing the respective
$\chi^2$-function. It is noteworthy that we do not aim to pinpoint
the right dark energy candidate among many competing models, but our
goal is to propose an effective and sensitive probe for testing the
validity of the concordance $\Lambda$CDM model. One possible
controversy here is, whether $E(z=0)=1$ is still valid for the case
of $n<0$, due to the divergence of the $Om(z)$ parametrization
proposed above. In fact, because the term of
$\left[(1+z)^3-1\right]$ in Eq.~\ref{Eom} approaches zero at $z=0$,
the convergent result of $E(z=0)=1$ in this case will be naturally
recovered, which can be clearly seen from the enlarged subplot in
Fig.~\ref{show_Ez}. More importantly, the $Om(z)$ parametrization
with different slope parameters also agrees very well with the
evolution of $E(z)$ reconstructed by $H(z)$ data with GP.

In this paper we use the latest $H(z)$ data set including 41
data points to place constrain on the $Om(z)$ parametrization
proposed above. In general, the measurement of $H(z)$ could be
obtained by two different techniques: galaxy differential age, also
known as cosmic chronometer (hereafter CC $H(z)$) and radial BAO
size methods (hereafter BAO $H(z)$) (\citealt{zhang2011constraints}). The latest 41 $H(z)$ data set including 31
CC $H(z)$ data and 10 BAO $H(z)$ data is compiled in Table~\ref{tab:
Hz}. Moreover, the Hubble function $H(z)$ should be normalized to
the dimensionless Hubble parameter $E(z)=H(z)/H_0$, whose
uncertainty could be obtained through
\begin{equation}
\sigma^2_E=\frac{\sigma^2_H}{H_0^2}+\frac{H^2}{H_0^4}\sigma^2_{H_0},
\end{equation}
where $\sigma_{H}$ and $\sigma_{H_0}$ are the uncertainty of $H(z)$
and $H_0$, respectively. In this work estimate the parameters by minimizing the $\chi^2 -$ function defined as
\begin{equation}
\chi^2_{H}({z,\textbf{p}})=\sum_{i=1}^{41}\frac{\left[E_{th}(z_i,\textbf{p})-E_{obs}(z_i)\right]^2}{\sigma_E(z_i)^2},
\end{equation}
where \textbf{p} denotes the $Om(z)$ parameters, $E_{th}$ and
$E_{obs}$ respectively stand for the theoretical and observed value
of the dimensionless Hubble parameter.

\begin{table}
\centering
\begin{tabular}{cccc}
\hline \hline
 \emph{z} & $H(z)$ & Method & Ref. \\
 & (km $\rm s^{-1}$ $\rm Mpc^{-1}$) &  & \\
\hline
0.09    &   $   69  \pm 12  $   & I &   \citealt{jimenez2003constraints} \\
\hline
0.17    &   $   83  \pm 8   $   &I &    \\
0.27    &   $   77  \pm 14  $   &I &    \\
0.4 &   $   95  \pm 17  $   &I &     \\
0.9 &   $   117 \pm 23  $   &I &    \citealt{simon2005improved} \\
1.3 &   $   168 \pm 17  $   &I &    \\
1.43    &   $   177 \pm 18  $   &I &    \\
1.53    &   $   140 \pm 14  $   &I &    \\
1.75    &   $   202 \pm 40  $   &I &    \\
\hline
0.48    &   $   97  \pm 62  $   &I &    \citealt{stern2010cosmic} \\
0.88    &   $   90  \pm 40  $   &I &    \\
\hline
0.35    &   $   82.1    \pm 4.9 $   &I &    \citealt{chuang2012measurements} \\
\hline
0.179   &   $   75  \pm 4   $   &I &    \\
0.199   &   $   75  \pm 5   $   &I &    \\
0.352   &   $   83  \pm 14  $   &I &    \\
0.593   &   $   104 \pm 13  $   &I &    \citealt{moresco2012new} \\
0.68    &   $   92  \pm 8   $   &I &    \\
0.781   &   $   105 \pm 12  $   &I &    \\
0.875   &   $   125 \pm 17  $   &I &    \\
1.037   &   $   154 \pm 20  $   &I &    \\
\hline
0.07    &   $   69  \pm 19.6    $   &I &    \\
0.12    &   $   68.6    \pm 26.2    $   &I &    \citealt{zhang2014four} \\
0.2 &   $   72.9    \pm 29.6    $   &I &    \\
0.28    &   $   88.8    \pm 36.6    $   &I &    \\
\hline
1.363   &   $   160 \pm 33.6    $   &I &    \citealt{moresco2015raising} \\
1.965   &   $   186.5   \pm 50.4    $   &I &    \\
\hline
0.3802  &   $   83  \pm 13.5    $   &I &    \\
0.4004  &   $   77  \pm 10.2    $   &I &    \\
0.4247  &   $   87.1    \pm 11.2    $   &I &    \citealt{moresco20166} \\
0.4497  &   $   92.8    \pm 12.9    $   &I &    \\
0.4783  &   $   80.9    \pm 9   $   &I &    \\
\hline
0.24    &   $   79.69   \pm 2.65    $   &II &   \citealt{gaztanaga2009clustering} \\
0.43    &   $   86.45   \pm 3.68    $   &II &   \\
\hline
0.44    &   $   82.6    \pm 7.8 $   &II &   \\
0.6 &   $   87.9    \pm 6.1 $   &II &   \citealt{blake2012wigglez}\\
0.73    &   $   97.3    \pm 7   $   &II &   \\
\hline
0.35    &   $   84.4    \pm 7   $   &II &   \citealt{xu2013measuring}\\
\hline
0.57    &   $   92.4    \pm 4.5 $   &II &   \citealt{samushia2013clustering}\\
\hline
2.3 &   $   224 \pm 8   $   &II &    \citealt{delubac2013baryon}\\
\hline
2.34    &   $   222 \pm 7   $   &II &   \citealt{delubac2015baryon} \\
\hline
2.36    &   $   226 \pm 8   $   &II &   \citealt{font2014quasar}\\
\hline \hline
\end{tabular}
\caption{\label{tab: Hz} The latest $H(z)$ measurements including 31
data points from the galaxy differential age method (I) and 10 data
points from the radial BAO size method (II)}. \label{table1}
\end{table}

\begin{table}
\centering
\begin{tabular}{l|l|l|ll}
\hline \hline
$H_0$ priors & $\alpha$ & $n$ &$\Lambda$CDM ($\alpha$, $n$)\\
\hline
Planck 2013 & $\alpha=0.268^{+0.02}_{-0.02}$  & $n=-0.172^{+0.12}_{-0.114}$  & (0.315, 0)\\
\hline
WMAP9 & $\alpha=0.268^{+0.026}_{-0.024}$ & $n=-0.021^{+0.151}_{-0.142}$ & (0.279, 0)\\
\hline
Riess (2016) & $\alpha=0.196^{+0.034}_{-0.032}$ & $n=0.162^{+0.163}_{-0.147}$ & (-, 0)\\
\hline \hline
\end{tabular}
\caption{\label{OHD} The best-fit values of the $Om(z)$ parameters
derived from the $H(z)$ data with different $H_0$ priors. The
corresponding values for $\Lambda$CDM are also presented for
comparison. }
\end{table}

\section{Results and discussion} \label{results}

\begin{figure}
\centering
\includegraphics[width=8cm,height=8cm]{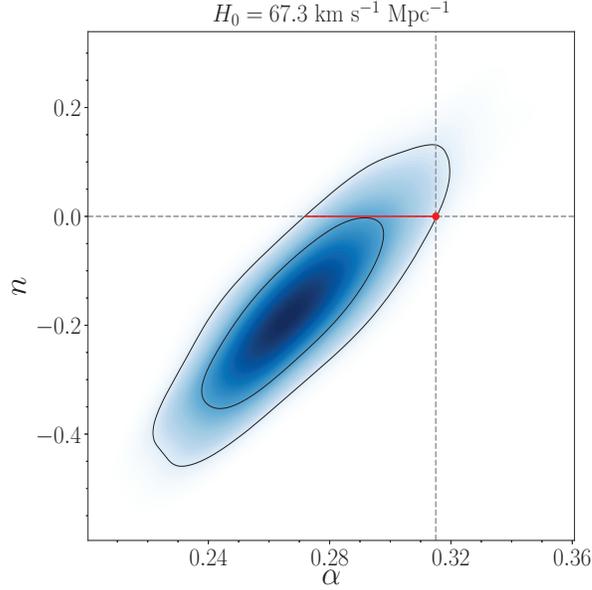}
\caption{ The $68.3\%$ and $95.4\%$ confidence regions in the
$\alpha-n$ parameter space for the $Om(z)$ parameterization
constrained from $H(z)$ data (with the prior of $H_0=67.3\pm 1.2$
km/s/Mpc from Planck). The red point represents the $\Lambda$CDM
model ($n=0.0$, $\Omega_{m0}=0.315$).}\label{oma1}
\end{figure}

\begin{figure}
\centering
\includegraphics[width=8cm,height=8cm]{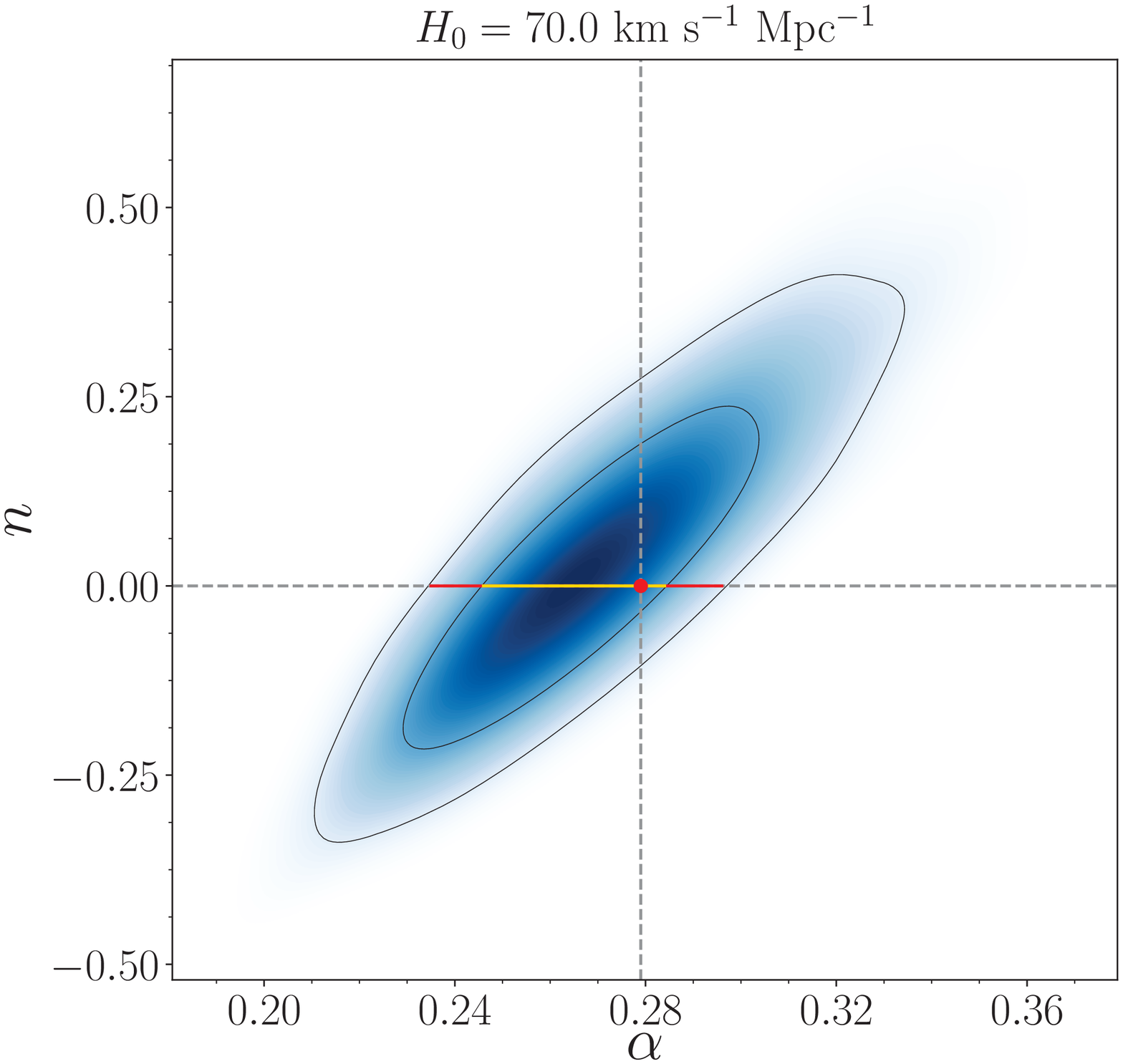}
\caption{The $68.3\%$ and $95.4\%$ confidence regions in the
$\alpha-n$ parameter space for the $Om(z)$ parameterization of
$Om(z)$ constrained from $H(z)$ data (with the prior of
$H_0=70.0\pm2.2$ km/s/Mpc from WMAP9). The red point represents
$\Lambda$CDM ($n=0.0$, $\Omega_{m0}=0.279$).}\label{oma2}
\end{figure}

\begin{figure}
\centering
\includegraphics[width=8cm,height=8cm]{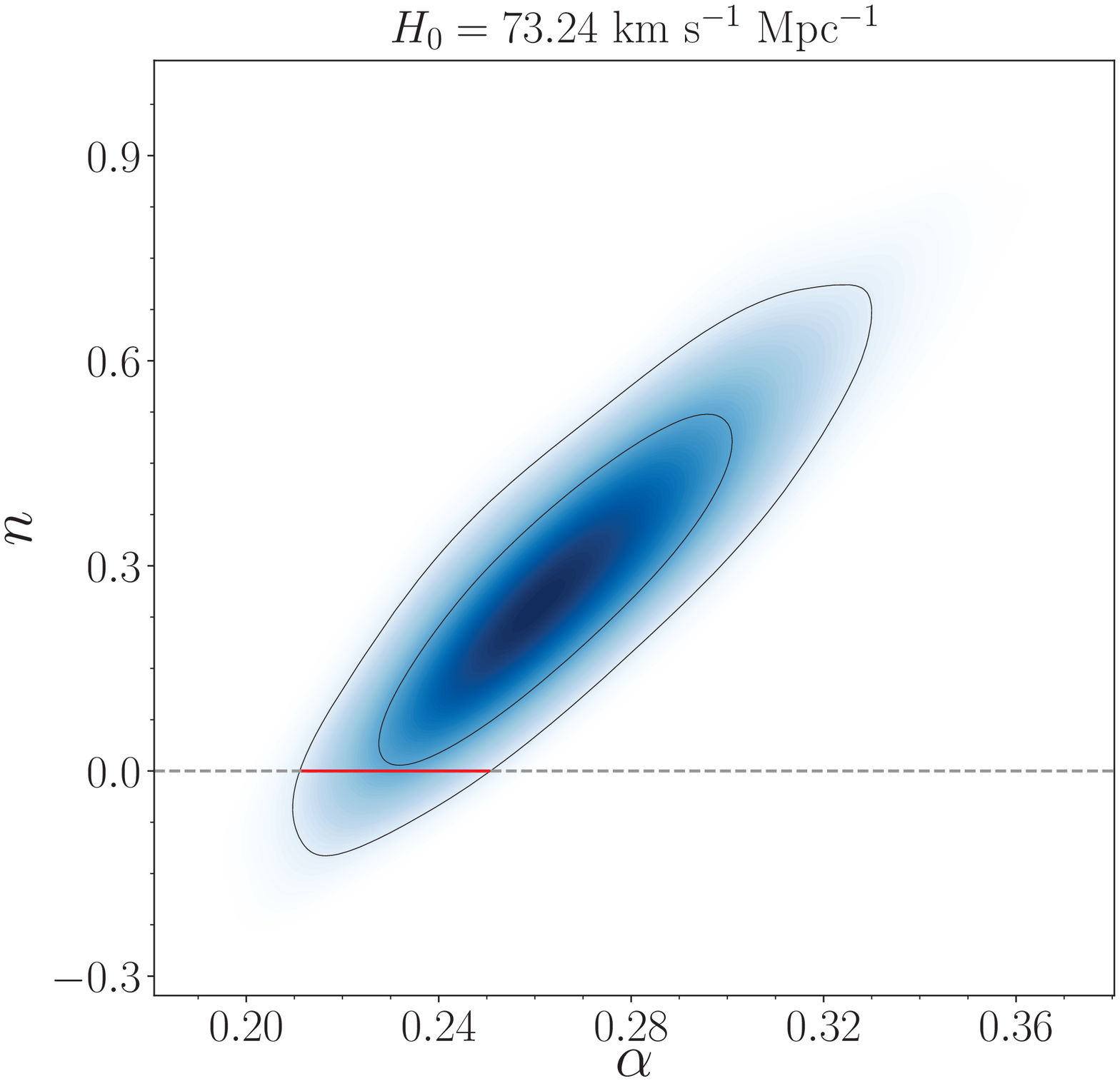}
\caption{ The $68.3\%$ and $95.4\%$ confidence regions in the
$\alpha-n$ parameter space for the $Om(z)$ parameterization of
$Om(z)$ constrained from $H(z)$ data (with the prior of
$H_0=73.24\pm1.74$ km/s/Mpc from \citet{riess20162}. The horizontal
black line denotes $\Lambda$CDM ($n=0.0$). }\label{oma3}
\end{figure}

We remark here that, as the benchmark of whole $H(z)$ data set, that the
influence of the Hubble constant value on the test of the $Om(z)$ parameter
should be taken into account.
Therefore, three recent measurements of
$H_{0}=67.3\pm1.2$ km $\rm s^{-1}$ $\rm Mpc^{-1}$ with $1.8\%$
uncertainty (\citealt{ade2014planck}), $H_{0}=70.0\pm2.2$ km $\rm
s^{-1}$ $\rm Mpc^{-1}$ with $3.1\%$ uncertainty
(\citealt{hinshaw2013nine}) and $H_{0}=73.24\pm1.74$ km $\rm s^{-1}$
$\rm Mpc^{-1}$ with $2.4\%$ uncertainty (\citealt{riess20162}) are
respectively used in our analysis. The
best-fit parameters (with 1$\sigma$ uncertainties) for these three
priors are presented in Table \ref{OHD}.

\subsection{Comparison with \textit{Planck} 2013 results}

To start with, we determine the best-fit $Om(z)$ parameters by
applying the Markov Chain Monte Carlo method to find the maximum likelihood
based on the $\chi^2$ - function. Then we compare the results with
cosmological parameters (\citealt{ade2014planck}) obtained by \textit{Planck} 2013 .
Adopting the Hubble
constant prior $H_0=67.3\pm 1.2$ km/s/Mpc to
the $H(z)$ data, we obtain the best-fit value of the parameters
$\alpha=0.268^{+0.02}_{-0.02}$ and $n=-0.172^{+0.12}_{-0.114}$ at
68:3\% confidence level. The value of the slope parameter $n$ is
obviously smaller than zero at $68\%$ confidence level (CL), which
implies that quintessence may be a good candidate of dark energy
as suggested by the $Om(z)$ parametrization. The marginalized 2D
confidence contours of $\alpha-n$ are shown in Fig.~2, in which the
$\Lambda$CDM model ($n=0.0$ and $\Omega_{m0}=0.315$) characterized
by \textit{Planck} 2013 data is also added for comparison. The deviation from
$\Lambda$CDM at $2\sigma$ confidence region strongly indicates a
tension between the current $H(z)$ data and $\Lambda$CDM, which
confirms the conclusion obtained in the previous works
(\citealt{sahni2014model,ding2015there,2016ApJ...825...17Z}).

\subsection{Comparison with WMAP9 results}

In the second case, we adopt the prior of $H_0=70.0\pm2.2$ km/s/Mpc
from WMAP9 results (\citealt{hinshaw2013nine}) to constrain the
parametrization of $Om(z)$. By minimizing the $\chi^2$, the
parameter $n$ implied by our statistical analysis gives
$n=-0.021^{+0.151}_{-0.142}$, which indicates that there is no
deviation from the $\Lambda$CDM scenario. Moreover, the best fit
obtained for the dark energy parameter is
$\alpha=0.268^{+0.026}_{-0.024}$ (68.3\% confidence level), which is
in perfect agreement with the matter density $\Omega_{m0}=0.279$
given by WMAP9. As can be seen from Fig.~\ref{oma2}, the discrepancy
between $H(z)$ data and $\Lambda$CDM determined by \textit{Planck} 2013 data has
gone, i.e., the observations of Hubble parameter still support the
existence of cosmological constant in the framework of this $Om(z)$
parametrization.

Obviously, the same $H(z)$ data set corresponding to different
values of $H_0$ and $\Omega_{m0}$ from Planck 2013 and WMAP9 gives
very different conclusions. Concerning the previous works
(\citealt{sahni2014model,ding2015there,2016ApJ...825...17Z}),
their estimates of $Omh^2$ are compared with $\Omega_{m0}h^2$, the
combination of $\Omega_{m0}$ and $H_0$ ($h=H_0/100$ km/s/Mpc)
determined by \textit{Planck} observations. Therefore, it is hard to tell
the source of the tension between $H(z)$ data and
$\Lambda$CDM. However, in our method, the impact of $H_0$ and
$\Omega_{m0}$ on the final conclusion could be separately discussed.
From the above analysis, we may conclude that the value of $H_0$ is
the most influential factor in performing consistency test of the
$\Lambda$CDM with $H(z)$ data. For instance, in the case of
$H_0=67.3\pm 1.2$ km/s/Mpc from \textit{Planck} 2013 data, $\Lambda$CDM with any
value of $\Omega_{m0}$ is ruled out at $68.3\%$ confidence level.
However, with the prior of $H_0=70.0\pm2.2$ km/s/Mpc from WMAP9, the
$H(z)$ data exhibits very good consistency with the concordance
cosmological constant model.

\subsection{Comparison with Riess (2016) results}

Considering the significant influence of $H_0$, in the final case a
local determination of $H_0=73.24\pm1.74$ km/s/Mpc with $2.4\%$
uncertainty from \citet{riess20162} can be taken to perform
consistency test. We show the contours constrained from the
statistical analysis in Fig.~\ref{oma3} and the best fit is
$\alpha=0.196^{+0.034}_{-0.032}$ and $n=0.162^{+0.163}_{-0.147}$.
($68.3\%$ confidence level). Different from the first case based on
\textit{Planck} measurements, a positive value for the slope parameter, which
corresponds to a phantom cosmology is favored in the framework of
this $Om(z)$ parametrization. Moreover, because this measurement of
$H_0$ is a local determination obtained in a cosmology-independent
method, we may comment on the value of matter density in
the framework of $\Lambda$CDM, i.e., at the $95.4\%$ confidence
level the range of $\Omega_{m0}$ is restricted to ($0.2118,0.2504$)
with the current $H(z)$ data, which is generally lower than the
value given by most of other types of cosmological observations.
Therefore, the measurement of $H_0$ from \citet{riess20162} will
significantly affect our understanding of $\Lambda$CDM and thus the
components in the Universe.

\begin{table}
\centering
\begin{tabular}{l|l|l|l}
\hline \hline
&$H_0$ & $\alpha$ & $n$ \\
\hline
JLA & $67.63^{+4.06}_{-2.18}$ & $0.292^{+0.097}_{-0.075}$ & $-0.014^{+0.240}_{-0.250}$ \\
\hline
$H(z)$ &$67.799^{+5.67}_{-13.29}$ & $0.276^{+0.070}_{-0.026}$& $-0.164^{+0.434}_{-0.545}$\\
\hline
$H(z)$+JLA &$68.81^{+1.50}_{-1.49}$ & $0.262^{+0.020}_{-0.018}$& $-0.096^{+0.105}_{-0.098}$\\
\hline \hline
\end{tabular}
\caption{\label{OHD+JLA} The best-fit values (with the 1$\sigma$
uncertainties) of the Hubble constant and the $Om(z)$ parameters
with different data combinations ($H(z)$, JLA and $H(z)$+JLA).}
\end{table}

\section{Constraints from JLA SN Ia sample} \label{JLA}

As mentioned above, the parametrization of $Om(z)$ proposed in this
paper makes it possible to perform consistency test of the
$\Lambda$CDM with other astronomical observations. More importantly,
previous literature have examined the role of $H(z)$ and SN Ia data
in cosmological constraints, and found that the they could play a
similar role in constraining the cosmological parameters
(\citealt{cao2010testing,cao2011interaction,cao2014testing}). Therefore,
we turn to the joint light-curve analysis (JLA) sample of 740 SNe Ia
data (\citealt{betoule2014improved}). For the JLA data, the observed
distance modulus is given by
\begin{equation}
\mu_{\mathrm{SN}}=m_{B}^{*}+\alpha\cdot x_1-\beta\cdot c-M_B
\end{equation}
where $m_{B}^{*}$ is the rest frame \textit{B}-band peak magnitude,
$x_1$ and $c$ are the time stretching of light-curve and the
supernova color at maximum brightness respectively, which are the
three parameters of light curve fitted by SALT2
(\citealt{guy2007salt2}). Moreover, the parameter $M_B$ describes the
absolute \textit{B}-band magnitude, whose value is assumed to be
dependent on the host stellar mass ($\textit{M}_{\mathrm{stellar}}$)
by a simple step function (\citealt{betoule2014improved})
\begin{equation}
M_B=\begin{cases}
M_B^1 &\text{for $M_{\mathrm{stellar}}<10^{10}M_{\odot}$}, \\
M_B^1+\Delta_M & \text{otherwise}.
\end{cases}
\end{equation}
Therefore, there are four nuisance parameters ($\alpha$, $\beta$,
$M_B^1$ and $\Delta_M$) to be fitted along with the $Om(z)$
parameters.

On the other hand, the theoretical distance modulus
$\mu_{\mathrm{th}}$ is expressed as $\mu_{\mathrm{th}}\equiv
5\log\left( D_L(z)/Mpc\right)+25$, where $D_L(z)$ is the luminosity
distance. Thus, the $\chi^2$ for the JLA sample is constructed as
\begin{equation}
\chi^2_{\mathrm{JLA}}=\Delta \mu^{T} \cdot \mathbf{Cov}^{-1} \cdot \Delta \mu,
\end{equation}
where $\Delta \mu=\mu_{\mathrm{SN}}(\alpha,~ \beta,~ M_B^1,~
\Delta_M;~ z)-\mu_{\mathrm{th}}(z)$ and $\mathbf{Cov}$ is the total
covariance matrix defined as
\begin{equation}
\mathbf{Cov}=\mathbf{D}_{\mathrm{stat}}+\mathbf{C}_{\mathrm{stat}}+\mathbf{C}_{\mathrm{sys}}.
\end{equation}
Here $\mathbf{D}_{\mathrm{stat}}$ corresponds to the diagonal part
of the statistical uncertainty, while $\mathbf{C}_{\mathrm{stat}}$
and $\mathbf{C}_{\mathrm{sys}}$ denote the statistical and
systematic covariance matrices, respectively. The details of the
covariance matrix $\mathbf{Cov}$ including its construction can be
found in \citet{betoule2014improved}. Considering the significance
of Hubble constant in testing $\Lambda$CDM, in this section we will
treat $H_0$ as a free parameter in the $\chi^2$-minimization
procedure. Thus there are four nuisance parameters plus tree
parameters ($H_0,~ \alpha,~ n$) referring to the parametrization of
$Om(z)$ that we are interested in.

\begin{figure*}
\centering
\includegraphics[width=10cm,height=10cm]{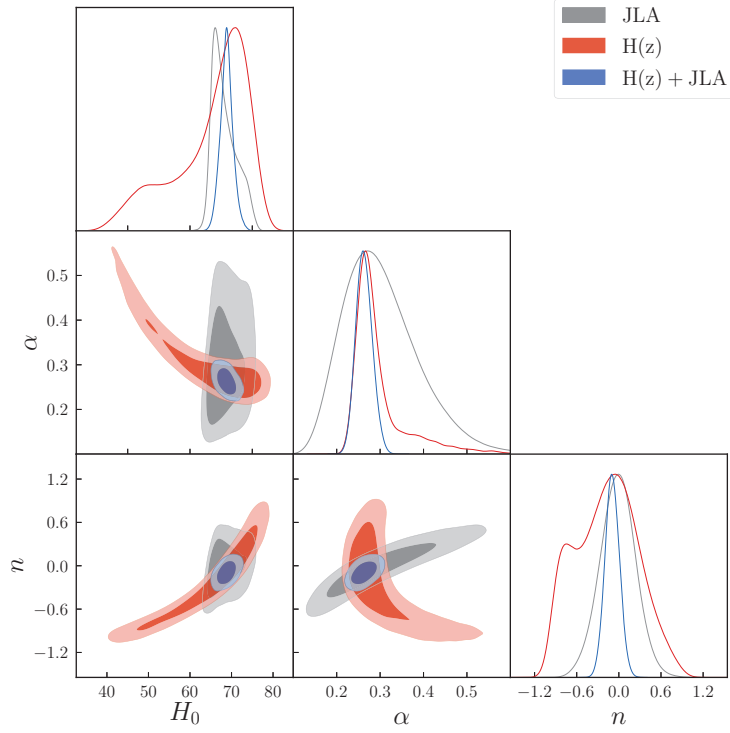}
\caption{ The $68.3\%$ and $95.4\%$ confidence regions in the
$H_0-\alpha-n$ parameter space derived from JLA, $H(z)$ and
$H(z)$+JLA.}\label{SNom}
\end{figure*}

In order to break the strong degeneracy between parameters, we also
perform a joint statistical analysis with use the JLA data and the
Hubble parameter measurements to constrain the parametrization of
$Om(z)$. The total $\chi^2$ with the combined data set of JLA and
$H(z)$ can be given by
\begin{equation}
\chi^2_{\mathrm{tot}}=\chi^2_{\mathrm{H}}+\chi^2_{\mathrm{JLA}}.
\end{equation}
The best-fit parameters (with 1$\sigma$ uncertainties) for different
data sets are presented in Table \ref{OHD+JLA}. The marginalized 2D
confidence contours of parameters ($\alpha$ and $n$, $\alpha$ and
$H_0$, and $n$ and $H_0$) are shown in Fig.~\ref{SNom}. It is
apparent that the principal axes of confidence regions obtained with
$H(z)$ data and JLA data intersect, which implies that the joint
analysis with $H(z)$ and JLA could effectively break the strong
degeneracy between parameters and thus provide a more stringent
constraint on the three parameters. On the one hand, although the
best-fit $Om(z)$ slope parameter is slightly smaller than zero,
which suggests that the current observational data tend to support a
quintessence cosmology, the $\Lambda$CDM model ($n=0$) is still
included within $1\sigma$ confidence region. On the other hand, the
best-fit Hubble constant from the combination of $H(z)$+JLA
($H_0=68.81^{+1.50}_{-1.49}$ km/s/Mpc) is consistent with the
results derived by both \textit{Planck} 2013 and WMAP9, which is
significantly different from the recent local measurement by
\citet{riess20162}.

\section{Conclusions and Discussions} \label{conclusion}

An important issue in modern cosmology is whether the EoS of dark
energy is a constant or varying with time. Based on an effective
diagnostic $Om(z)$ and its improved version $Om(z_1,z_2)$ and
$Omh^2(z_1,z_2)$, many recent works have performed a null test of
$\Lambda$CDM determined by \textit{Planck} observations, which implies that
the $\Lambda$CDM model may not be the best scenario of our universe,
or dark energy does not exist in the form of the cosmological
constant. In this paper, we have proposed a parametrization of
$Om(z)$ to investigate the validity of $\Lambda$CDM, which
successfully clarify the impact of $H_0$ and $\Omega_{m0}$ on the final conclusion. With three different priors of the Hubble
constant $H_0$, the latest $H(z)$ data is used to set constraint on
the $Om(z)$ parameters of interest. Our results showed that the
value of $H_0$ plays a very import role in the consistency test of
$\Lambda$CDM. Here we summarize our main conclusions in more detail:
\begin{itemize}
  \item Adopting the Hubble constant prior $H_0=67.3\pm 1.2$ km/s/Mpc
(\citealt{ade2014planck}) to the $H(z)$ data, we find the value of the
slope parameter $n$ smaller than zero at $68\%$
confidence level, which implies that quintessence may be a good
candidate of dark energy according to this $Om(z)$
parametrization. The deviation from $\Lambda$CDM at $2\sigma$
confidence region strongly indicates a tension between the current
$H(z)$ data and $\Lambda$CDM, which confirms the conclusion obtained
in the previous works.
\item With the prior of $H_0=70.0\pm2.2$
km/s/Mpc from WMAP9 results, the discrepancy between $H(z)$ data and
$\Lambda$CDM disappeared, i.e., the
data analyzed in the framework of this $Om(z)$
parametrization still support the cosmological constant scenario.
  \item In the third case with the local determination of
$H_0=73.24\pm1.74$ km/s/Mpc from \citet{riess20162}, a positive
value for the slope parameter, which corresponds to a phantom
cosmology is strongly favored by the current $H(z)$ data. Moreover,
at the $95.4\%$ confidence level the range of matter density is
restricted to $\Omega_{m0}=(0.2118,0.2504)$, which is generally
lower than the value given by most of other types of cosmological
observations.
\end{itemize}

Moreover, the parametrization of $Om(z)$ makes it possible to
perform consistency test of the $\Lambda$CDM with other astronomical
observations. We studied the constraining power of the JLA sample of
740 SNe Ia data (\citealt{betoule2014improved}) and its combination with
the Hubble parameter measurements on the parametrization of $Om(z)$.
Here we summarize our main conclusions in more detail:
\begin{itemize}
  \item Although the best-fit $Om(z)$ slope parameter is
slightly smaller than zero, which suggests that the current
observational data tend to support a quintessence cosmology, the
$\Lambda$CDM model ($n=0$) is still included within $1\sigma$
confidence region.
\item The best-fit Hubble constant
from the combination of $H(z)$+JLA ($H_0=68.81^{+1.50}_{-1.49}$
km/s/Mpc) is well consistent with the results derived both by \textit{Planck}
2013 and WMAP9, which is significantly different from the recent
local measurement by \citet{riess20162}.
\end{itemize}

As a final remark, the parametrization of $Om(z)$ proposed in this
paper has opened a robust window for testing the validity of the
concordance $\Lambda$CDM cosmology and suggesting the other possible
dynamical dark energy models. However, more precise model selection
still remains a difficult task with the current accuracy of the data
and the important role played by the Hubble constant. We hope that
future data concerning strong gravitational lensing observations
(\citealt{cao2011testing,cao2011constraints,cao2012constraints,cao2012testing,cao2015cosmology}),
high-redshift SN Ia from SDSS-II and SNLS collaborations
(\citealt{betoule2014improved}), ultra-compact structure in
high-redshift radio quasars (\citealt{cao2017a}), and weak lensing
surveys combined with CMB measurements (\citealt{ade2016planck}) will
lead to a substantial progress in this respect.

\normalem
\begin{acknowledgements}

This work was supported by National Key R\&D Program of China No.
2017YFA0402600, the National Basic Science Program (Project 973) of
China under (Grant No. 2014CB845800), the National Natural Science
Foundation of China under Grants Nos. 11503001, 11690023, 11373014,
and 11633001, the Strategic Priority Research Program of the Chinese
Academy of Sciences, Grant No. XDB23000000, the Interdiscipline
Research Funds of Beijing Normal University, and the Opening Project
of Key Laboratory of Computational Astrophysics, National
Astronomical Observatories, Chinese Academy of Sciences. J.-Z. Qi
was supported by China Postdoctoral Science Foundation under Grant
No. 2017M620661. This research was also partly supported by the
Poland-China Scientific \& Technological Cooperation Committee
Project No. 35-4. M.B. was supported by Foreign Talent Introducing
Project and Special Fund Support of Foreign Knowledge Introducing
Project in China.

\end{acknowledgements}

\bibliographystyle{raa}
\bibliography{ombib}

\end{document}